\documentclass[onecollarge,natbib]{svjour2}
\bibpunct{[}{]}{;}{n}{}{,} 
\smartqed  
\usepackage{graphicx}
\usepackage{mathptmx}      
\usepackage{latexsym}
\usepackage{amsmath}
\usepackage{amssymb}

\newcommand{\bbR}{{\mathbb R}}

\newcommand{\cO}{{\mathcal O}}
\newcommand{\sE}{{\sf E}}
\newcommand{\fH}{\mathfrak{H}}
\newcommand{\fP}{\mathfrak{P}}
\newcommand{\fQ}{\mathfrak{Q}}
\newcommand{\dist}{\mathop{\rm dist}}

\newcommand{\Ran}{\mathop{\mathrm{Ran}}}
\DeclareMathOperator{\spec}{spec}

\journalname{}
\begin{document}

\title{On applying the subspace perturbation theory\\
to few-body Hamiltonians\thanks{Based on a talk presented at the 22nd
European Conference on Few-Body Problems in Physics (September
9--13, 2013, Cracow, Poland). The paper is to be published in
\textit{Few-Body Systems}, doi: 10.1007/s00601-013-0752-8. This work
was supported by the Heisenberg-Landau Program and by the Russian
Foundation for Basic Research.}}

\titlerunning{On applying the subspace perturbation theory}

\author{Alexander K. Motovilov}

\authorrunning{A.K.Motovilov} 

\institute{A.K.Motovilov \at
           Bogoliubov Laboratory of Theoretical Physics, JINR\\
           Joliot-Curie 6, 141980 Dubna, Moscow Region, Russia\\
              Tel.: +7-496-216-3355\\
              \email{motovilv@theor.jinr.ru}
}


\maketitle

\begin{abstract}
We present a selection of results on variation of the spectral
subspace of a Hermitian operator under a Hermitian perturbation and
show how these results may work for few-body Hamiltonians.

\keywords{Few-body problem \and Binding energy shift \and Variation of spectral subspace}
\end{abstract}

\section{Introduction}
\label{intro}

The subspace perturbation theory is a branch of the general theory
of linear operators (see, e.g. \cite{AkhiG,Kato}) that studies
variation of an invariant (in particular, spectral) subspace of an
operator under an additive perturbation. In this small survey
article we restrict the subject to Hermitian operators and follow
the geometric approach that originates in the works by Davis
\cite{D63,D65} and Davis and Kahan \cite{DK70}. Within the
Davis-Kahan approach, a bound on the variation of a spectral
subspace has usually the form of a trigonometric estimate involving
just two quantities: a norm of the perturbation operator and the
distance between the relevant spectral subsets. We present
only the estimates that are a priori in their nature and involve the
distance between spectral sets of the unperturbed operator (but not
of the perturbed one). The results valid for Hermitian
operators of any origin are collected in Section \ref{Sec2}. In
Section \ref{Sec-FB} we give several examples that illustrate the
meaning of the abstract results and show why these results might be
useful already in the study of a few-body bound-state problem.

Through the whole material we will use only the standard operator
norm. We recall that if $V$ is a bounded linear operator on a
Hilbert space $\fH$, its norm is given by
$\|V\|=\sup_{\|f\|=1} \bigl\|V|f\rangle\bigr\|$ where sup denotes
the least upper bound. Thus, for any {$|f\rangle\in\fH$} we have
$\bigl\|V|f\rangle\bigr\|\leq \|V\|\,\|f\|$. If {$V$} is a Hermitian
operator with $\min\bigl(\spec({V})\bigr)=m_V$ and
$\max\bigl(\spec({V})\bigr)=M_V$ then $\|V\|=\max\{|m_V|,|M_V|\}$.
In particular, if $V$ is separable of rank one, that is, if
{$V=\lambda|\phi\rangle\langle \phi|$} with a normalized
$|\phi\rangle\in\fH$ and $\lambda\in\bbR$, then $\|V\|=|\lambda|$.
Another example concerns the case where $\fH=L_2(\bbR)$ and
$V$ is a bounded local potential, i.e. $\langle
x|V|f\rangle=V(x)f(x)$ for any $|f\rangle\in\fH$, with $V(\cdot)$ a
bounded function on {$\bbR$}. In this case
{$\|V\|=\sup_{x\in\bbR}|V(x)|$}.

\section{The abstract problem setup and abstract results}
\label{Sec2}

Assume that $A$ is a Hermitian (or, equivalently, self-adjoint)
operator on a separable Hilbert space $\fH$. If $V$ is a bounded
Hermitian perturbation of $A$ then the spectrum, $\spec(H)$, of the
perturbed operator $H=A+V$ is confined in the closed
$\|V\|$-neighborhood $\cO_{\|V\|}\bigl(\spec(A)\bigr)$ of the
spectrum of $A$ (see, e.g., \cite{Kato}). Hence, if a part $\sigma$
of the spectrum of $A$ is isolated from its complement
$\Sigma=\spec(A)\setminus\sigma$, that is, if
\begin{equation}
\label{separIn}
d:=\dist(\sigma,\Sigma)>0,
\end{equation}
then the spectrum of $H$ is also divided into two disjoint
components,
\begin{equation}
\label{omOMd}
\omega=\spec(H)\cap\cO_{\|V\|}(\sigma)\quad\text{and}\quad
\Omega=\spec(H)\cap\cO_{\|V\|}(\Sigma),
\end{equation}
provided that
\begin{equation}
\label{Vd12In}
 \|V\|<\text{\small$\frac{1}{2}$}\,d.
\end{equation}
Under condition \eqref{Vd12In}, one interprets the separated
spectral components $\omega$ and $\Omega$ of the perturbed operator
$H$ as the results of the perturbation of the corresponding initial
disjoint spectral sets $\sigma$ and $\Sigma$.

The transformation of the spectral subspace of $A$ associated with
the spectral set $\sigma$ into the spectral subspace of $H$
associated with the spectral set $\omega$ may be studied in terms of
the difference $P-Q$ between the corresponding spectral projections
$P=\sE_A(\sigma)$ and $Q=\sE_H(\omega)$ of $A$ and $H$. Of
particular interest is the case where $\|P-Q\|<1$. In this case the
spectral projections $P$ and $Q$ are unitarily equivalent and the
perturbed spectral subspace $\fQ=\Ran(Q)$ may be viewed as obtained
by the direct rotation of the unperturbed spectral subspace
$\fP=\Ran(P)$ (see, e.g. \cite[Sections 3 and 4]{DK70}). Moreover,
the quantity
$$
\theta(\fP,\fQ)=\arcsin(\|P-Q\|)
$$
is used as a measure of this rotation. It is called the maximal
angle between the subspaces $\fP$ and $\fQ$. A short review of the
concept of the maximal angle can be found, e.g., in \cite[Section
2]{AM-CAOT-2013}; see also \cite{DK70,KMM5,MotSel,Se2013-1}.

Among various questions being answered within the subspace perturbation
theory, the first and rather basic question is on whether the
condition \eqref{Vd12In} is sufficient for the bound
\begin{equation}
\label{tlp2}
\theta(\fP,\fQ)<\text{\small$\frac{\pi}{2}$}
\end{equation}
to hold, or, in order to secure \eqref{tlp2}, one should impose a
stronger requirement $\|V\|<c\,d$ with some $c<\frac{1}{2}$. More
precisely, the question is as follows. \vspace*{-0.3em}
\begin{enumerate}
\item[(i)] What is the best possible constant
$c_*$ in the inequality $\|V\|<c_*\,d$ ensuring
the spectral subspace variation bound \eqref{tlp2}?
\end{enumerate}
\vspace*{-0.2em} \noindent Another, practically important question
concerns the maximal possible size of the subspace variation:
\vspace*{-0.2em}
\begin{enumerate}
\item[(ii)]  What function $M: [0,c_*)\mapsto \bigl[0,\frac{\pi}{2}\bigr)$
is best possible
in the bound
\begin{equation}
\label{Mquest}
\theta(\fP,\fQ)\leq M\left(\text{\small$\frac{\|V\|}{d}$}\right)\quad\text{for}\quad \|V\|<c_*\,d?
\end{equation}
\end{enumerate}
Both the constant $c_*$ and the estimating function $M$
are required to be universal in the sense that they should not
depend on the Hermitian operators $A$ and $V$ involved.

By now, the problems (i) and (ii) have been completely solved only
for the particular spectral dispositions where one of the sets
$\sigma$ and $\Sigma$ lies in a finite or infinite gap of the other
set, say, $\sigma$ lies in a gap of $\Sigma$. In this case
\begin{equation}
\label{sin2t}
c_*=\text{\small$\frac{1}{2}$}\quad\text{and}\quad M(x)=\text{\small$\frac{1}{2}$}\arcsin(2x).
\end{equation}
This is the essence of the Davis-Kahan $\sin2\theta$ theorem in
\cite{DK70}.

In the generic case (where no assumptions are done on the mutual
position of the sets $\sigma$ and $\Sigma$), the strongest known
answers to the questions (i) and (ii) are the recent ones given by
Seelmann \cite{Se2013-2}, within the approach developed in
\cite{AM-CAOT-2013}; see also the earlier works \cite{KMMa} and
\cite{MS2010}. In particular, \cite[Theorem 1]{Se2013-2} implies
that the generic optimal constant $c_*$ satisfies inequality $
c_*>0.454839. $ For the best upper estimate on the function $M$ in
the bound \eqref{Mquest} we also refer to \cite{Se2013-2}. Here we
only notice that for sure $\theta(\fP,\fQ)\leq
\frac{1}{2}\arcsin\frac{\pi\|V\|}{d}$ whenever
$\|V\|\leq\frac{1}{\pi}\,d$ (see \cite[Remark 4.4]{AM-CAOT-2013};
cf. \cite[Corollary 2]{Se2013-1}).

Now recall that, under the assumption \eqref{separIn}, a bounded
operator $V$ is said to be off-diagonal with respect to the
partition $\spec(A)=\sigma\cup\Sigma$ if it anticommutes with the
difference $J=P-P^\perp$ of the spectral projections
$P=\sE_A(\sigma)$ and $P^\perp=\sE_A(\Sigma)$, that is, if\,
$VJ=-JV$. The problems like (i) and (ii) have also been addressed in
the case of off-diagonal Hermitian perturbations. When considering
such a perturbation, one should first take into account that the
requirements ensuring the disjointness of the corresponding
perturbed spectral components $\omega$ and $\Omega$ originating in
$\sigma$ and $\Sigma$ are much weaker than condition \eqref{Vd12In}.
In particular, if the sets $\sigma$ and $\Sigma$ are subordinated,
say $\max(\sigma)<\min(\Sigma)$, then for any $\|V\|$ no spectrum of
$H=A+V$ enters the open interval between $\max(\sigma)$ and
$\min(\Sigma)$ (see, e.g., \cite[Remark 2.5.19]{Tretter2008}). In such
a case the maximal angle between the unperturbed and perturbed
spectral subspaces admits a sharp bound of the form \eqref{Mquest}
with $\text{$\,M(x)=\text{$\frac{1}{2}$}\arctan(2x)$,
$\,x\in[0,\infty)$.}$ This is the consequence of the celebrated
Davis-Kahan $\tan2\theta$ theorem \cite{DK70} (also, cf. the
extensions of the $\tan2\theta$ theorem in
\cite{GKMV2013,KMM5,MotSel}). Furthermore, if it is known that the
set $\sigma$ lies in a finite gap of the set $\Sigma$ then the
disjointness of the perturbed spectral components $\omega$ and
$\Omega$ is ensured by the (sharp) condition $\|V\|<\sqrt{2}\,d$.
The same condition is optimal for the bound \eqref{tlp2} to hold.
Both these results have been established in \cite{KMM3}. The
explicit expression for the best possible function $M$ in the
corresponding estimate \eqref{Mquest}, $M(x)=\arctan x$,
$x\in[0,\sqrt{2})$, was found in \cite{AM-IEOT-2012,MotSel}. As for
the generic spectral disposition with no restrictions on the mutual
position of $\sigma$ and $\Sigma$,  the condition
$\|V\|<\frac{\sqrt{3}}{2}d$ has been proven to be optimal in order
to guarantee that the gaps between $\sigma$ and $\Sigma$ do not
close under an off-diagonal $V$ and, thus, that
$\dist(\omega,\Omega)>0$. The proof was first given in
\cite[Theorem~1]{KMM4} for bounded $A$ and then in \cite[Proposition
2.5.22]{Tretter2008} for unbounded $A$. The best published lower
bound  $c_*> 0.675989$ for the generic optimal constant $c_*$ in the
off-diagonal case has been established in \cite{MS2010}. Paper
\cite{MS2010} also contains the strongest known upper estimate for
the optimal function $M$ in the corresponding bound \eqref{Mquest}
(see \cite[Theorem 6.2 and Remark 6.3]{MS2010}).

\section{Applications to few-body problems}
\label{Sec-FB}

Throughout this section we suppose that the ``unperturbed''
Hamiltonian $A$ has the form $A=H_0+V_0$ where $H_0$ stands for the
kinetic energy operator of an $N$-particle system in the c.m. frame
and the potential $V_0$ includes only a part of all interactions
that are present in the system (say, only two-body forces if $N=3$).
The perturbation $V$ is assumed to describe the remaining part of
the interactions (say, three-body forces for $N=3$; instead, if all
the interparticle interactions are already included in $V_0$, it may
only describe the effect of external fields). We consider the case
where $V$ is a bounded operator. Surely, both $A$ and $V$ are
assumed to be Hermitian. In order to apply to $H=A+V$ the abstract
results mentioned in the previous section, one only needs to know
the norm of the perturbation $V$ and a few very basic things on the
spectrum of the operator~$A$.

We start our discussion with the simplest example illustrating
the $\sin2\theta$ and $\tan2\theta$ theorems from \cite{DK70}.
\smallskip

\textit{Example 3.1} Suppose that {$E_0$} is the ground-state (g.s.)
energy of the Hamiltonian $A$. Assume, in addition, that the
eigenvalue $E_0$ is simple (this is typical for a ground state) and
let $|\psi_0\rangle$ be the normalized g.s. wave function, i.e.
$A|\psi_0\rangle=E_0|\psi_0\rangle$, $\|\psi_0\|=1$. Set
$\sigma=\{E_0\}$, $\Sigma=\spec(A)\setminus\{E_0\}$ and
$d=\dist(\sigma,\Sigma)=\min(\Sigma)-E_0$ (we remark that the set
$\Sigma$ is definitely not empty since it should contain at least
the essential spectrum of $A$). If the norm of $V$ is such that the
condition \eqref{Vd12In} holds, then the g.s. energy $E'_0$ of the
total Hamiltonian $H=A+V$ is also a simple eigenvalue with a g.s.
vector $|\psi'_0\rangle$, $\|\psi'_0\|=1$. The g.s. energy $E'_0$
lies in the closed $\|V\|$-neighborhood of the g.s. energy $E_0$,
that is, $|E_0-E'_0|\leq\|V\|$. The corresponding spectral
projections $P=\sE_A(\sigma)$ and $Q=\sE_H(\omega)$ of $A$ and $H$
associated with the one-point spectral sets $\sigma=\{E_0\}$ and
$\omega=\{E'_0\}$ read as $P=|\psi_0\rangle\langle\psi_0|$ and
$Q=|\psi'_0\rangle\langle\psi'_0|$. One verifies by inspection that\,
$\arcsin\bigl(\|P-Q\|\bigr)=\arccos|\langle\psi_0|\psi'_0\rangle|$, which means
that the maximal angle\, $\theta(\fP,\fQ)$ \,between the one-dimensional
spectral subspaces $\fP=\Ran(P)=\mathop{\rm span}(|\psi_0\rangle)$
and $\fQ=\Ran(Q)=\mathop{\rm span}(|\psi'_0\rangle)$ is, of course,
nothing but the angle between the g.s. vectors
$|\psi_0\rangle$ and $|\psi'_0\rangle$. Then the Davis-Kahan
$\sin2\theta$ theorem (see \eqref{Mquest} and \eqref{sin2t}) implies
that\,
$\arccos|\langle\psi_0|\psi'_0\rangle|\leq\frac{1}{2}\arcsin\frac{2\|V\|}{d}.$
This bound on the rotation of the ground state under the
perturbation $V$ is sharp. In particular, it implies that under
condition \eqref{Vd12In} the angle between $|\psi_0\rangle$ and
$|\psi'_0\rangle$ can never exceed~$45^\circ$.

If, in addition, it is known that the perturbation $V$ is
off-diagonal with respect to the partition
$\spec(A)=\sigma\cup\Sigma$ then for any (arbitrarily large) {$\|V\|$}
no spectrum of $H$ enters the gap between the g.s. energy $E_0$ and
the remaining spectrum $\Sigma$ of $A$. Moreover, there are the
following sharp universal bounds for the perturbed g.s. energy $E'_0$:
$$
\mbox{$E_0-\epsilon_V\leq E'_0\leq E_0$},
$$
where
\begin{equation}
\label{rV}
\epsilon_V=\|V\|\tan\left(\text{\small$\frac{1}{2}$}\arctan\text{\small$\frac{2\|V\|}{d}$}\right)<\|V\|
\end{equation}
(see \cite[Lemma 1.1]{KMM4} and \cite[Proposition
2.5.21]{Tretter2008}).  At the same time, the Davis-Kahan
$\tan2\theta$ theorem \cite{DK70} implies that\,
$\arccos|\langle\psi_0|\psi'_0\rangle|\leq\frac{1}{2}\arctan\frac{2\|V\|}{d}<\frac{\pi}{4}.$

With a minimal change, the above consideration is extended to the
case where the initial spectral set $\sigma$ consists of the $n+1$
lowest binding energies $E_0\leq E_1\leq\ldots\leq E_n$, $n\geq 1$,
of $A$. We only want to underline that if $V$ is off-diagonal than
for any $\|V\|$ the whole perturbed spectral set $\omega$ of $H=A+V$
originating from $\sigma$ will necessarily lie in the interval
$[E_0-r_V,E_{n}]$ where the shift $r_V$ is given by \eqref{rV}, while the
interval $\bigl(E_n,\min(\Sigma)\bigr)$ will contain no spectrum of $H$.
Furthermore, the $\tan2\theta$-theorem-like estimates for the maximal angle
between the spectral subspaces $\fP=\Ran\bigl(\sE_A(\sigma)\bigr)$
and $\fQ=\Ran\bigl(\sE_H(\omega)\bigr)$ may be done even for some
unbounded $V$ (but, instead of $d$ and $\|V\|$, in terms of
quadratic forms involving $A$ and $V$), see \cite{GKMV2013,MotSel}.
\smallskip

Along with the $\sin2\theta$ theorem, our second example also illustrates
the $\tan\theta$ bound proven in \cite{AM-IEOT-2012,MotSel}.

\smallskip

\textit{Example 3.2} Assume that
${\sigma=\{E_{n+1},E_{n+2},\ldots,E_{n+k}\}},\, {n\geq 0,\,\, k\geq
1},$ is the set of consecutive binding energies of {$A$} and
$\Sigma=\spec(A)\setminus\sigma=\Sigma_-\cup\Sigma_+$ where
$\Sigma_-$ is the increasing sequence of the energy levels $E_0,
E_1,\ldots,E_n$ of $A$ that are lower than $\min(\sigma)$;
$\Sigma_+$ denotes the remaining part of the spectrum of $A$, i.e.
$\Sigma_+=\spec(A)\setminus(\sigma\cup\Sigma_-)$. Together with
\eqref{separIn}, this assumption means that the set $\sigma$ lies in
the finite gap $\bigl(\max(\Sigma_-),\min(\Sigma_+)\bigr)$ of the
set $\Sigma$. Under the single condition \eqref{Vd12In}, not much
can be said about the location of the perturbed spectral sets
$\omega$ and $\Omega$, except for \eqref{omOMd}, but the Davis-Kahan
$\sin2\theta$ theorem \cite{DK70} still well applies to this case
and, thus, we again have the bound
$\theta(\fP,\fQ)\leq\frac{1}{2}\arcsin\frac{2\|V\|}{d}<\frac{\pi}{4}$.

Much stronger conclusions can be done if the operator $V$ is
off-diagonal with respect to the partition
$\spec(A)=\sigma\cup\Sigma$. As it was already mentioned in Section
2, for off-diagonal $V$ the gap-non-closing condition reads as
$\|V\|<\sqrt{2}d$ (and even a weaker but more detail condition
$\|V\|<\sqrt{dD}$\, with\, $D=\min(\Sigma_+)-\max(\Sigma_-)$\, is
allowed, see \cite{KMM3,MotSel}). In this case the lower bound for
the spectrum of $H=A+V$ is $E_0-\epsilon_V$ where the maximal
possible energy shift $\epsilon_V$, $\epsilon_V<d$, is given again
by \eqref{rV}. Furthermore,
$\omega\subset[E_{n+1}-\epsilon_V,E_{n+k}+\epsilon_V]$ and the open
intervals $(E_n,E_{n+1}-\epsilon_V)$ and
$\bigl(E_{n+k}+\epsilon_V,\min(\Sigma_+)\bigr)$ contain no spectrum
of $H$. For tighter enclosures for $\omega$ and $\Omega$ involving,
say, the the gap length\, $D$\, we refer to
\cite{KMM3,KMM4,Tretter2008}. In the case under consideration, the
sharp bound for the rotation of the spectral subspace
$\fP=\Ran\bigl(\sE_A(\sigma)\bigr)$ is given by
$\theta(\fP,\fQ)\leq\arctan\text{$\frac{\|V\|}{d}$}<\arctan\sqrt{2}$
(see \cite[Theorem 1]{AM-IEOT-2012}; cf. \cite[Theorem 2]{MotSel}).
However, if the value of $D$\, is known and $\|V\|<\sqrt{dD}$, then a
more detail and stronger but still optimal estimate for
$\theta(\fP,\fQ)$ involving $D$ is available (see \cite[Theorem
4.1]{AM-IEOT-2012}).
\smallskip

Both Examples 3.1 and 3.2 show how one may obtain a bound on the
variation of the spectral subspace prior to any calculations with
the total Hamiltonian $H$. To perform this, only the knowledge of
the values of $d$ and $\|V\|$ is needed. Furthermore, if $V$ is
off-diagonal,   with just these two values one can also provide the
stronger estimates (via $\epsilon_V$) for the binding energy shifts.

{\small

\end{document}